\documentclass[a4paper,twoside,11pt]{article} % to use for all kind of papers but AACA

\usepackage[english]{babel} % 
\usepackage{lmodern}
\usepackage[T1]{fontenc}
\usepackage[latin9]{inputenc}	% Così le "è" vengono trasformate in \`e etc. etc.
\newlength{\defaultparindent}
\usepackage{latexsym}
\usepackage{amsmath}
\usepackage{amsfonts}
\usepackage{amsthm}
\usepackage{bbold}
\usepackage{multirow}
\usepackage{hyperref}
\hypersetup{colorlinks=true,citecolor=green,linkcolor= blue}
\usepackage{todonotes} % more appealing than \marginpar ??
\usepackage{soul} % to get overstriked text

\usepackage[arXiv]{optional} % Possible values of the option fixing the format: x, std, arXiv, JMP, JOPA, AACA and also note control: margin_notes, final_notes
\def\mynote{\todo} % \marginpar or \todo
\opt{AACA}{% only for submission to AACA -> \cal font is not defined....
\def\cal{\mathcal}
}

\opt{x,AACA}{% in all cases - but arXiv, std, JMP & JOPA - include my file of standard defs: mbh_defs_TeX
\input{mbh_defs_TeX}% My standard TeX definitions, original in Dictionaries & Gloassaries folder
}

\opt{arXiv,std,JMP,JOPA}{% only for arXiv, std, JMP & JOPA we need the definitions included in mbh_defs_TeX (that contains original versions)

\newtheorem{MS_Proposition}{Proposition}

\def\ie{i.e.\ }
\def\eg{e.g.\ }
\def\myisom{\cong} % this way isomorphism symbol definition can be easily changed; possiblities: \cong \simeq
\newcommand{\R}{\ensuremath{\mathbb{R}}} % good; MB XII 2005; needs \usepackage{bbold}
\newcommand{\C}{\ensuremath{\mathbb{C}}} % good; MB XII 2005; needs \usepackage{bbold}
\newcommand{\F}{\ensuremath{\mathbb{F}}} % good; MB XII 2005; needs \usepackage{bbold}
\newcommand{\Z}{\ensuremath{\mathbb{Z}}} % good; MB XII 2005; needs \usepackage{bbold}
\newcommand{\HH}{\ensuremath{\mathbb{H}}} % good; MB V 2015; \H already defined; needs \usepackage{bbold}
\newcommand{\Identity}{\ensuremath{\mathbb{1}}} % good; MB XII 2005; needs \usepackage{bbold}
\def\my_span#1{\mbox{Span}\left(#1\right)} % changed from 'span' since it interfered with \multicolumn{} MB X 2009
\def\mytrace#1{\mbox{Tr}\left(#1\right)} % if I call this my_trace it messes with my_span ??? MB VII 2012

\def\dotinformula{\;\; \mathrm{.}} % defines space + a full stop (in \rm font) to be placed at the end of a formula
\def\OO#1{\ensuremath{\mbox{O}\left(#1\right)}}

\def\Pin#1{\ensuremath{\mbox{Pin}\left(#1\right)}}

\def\End{\ensuremath{\mbox{End}}}
\newcommand{\comm}[2]{\ensuremath{\left[ #1, #2 \right]}}
\newcommand{\anticomm}[2]{\ensuremath{\left\{ #1, #2 \right\}}} % or \left[...\right]^+
\def\Cl{{\cal C}\ell}	% this is a def from Dilys Grilli @ ICTP in Paolo's paper, IV 2009, superseded by following myCl
\newcommand{\myCl}[3]{\ensuremath{{{\cal C}\ell} {\left( #1, #2 \right)}}}	% this is a lighter def of myCl without the scalar product g
\newcommand{\myClf}[3]{\ensuremath{{{\cal C}\ell}_{#3} {\left( #1, #2 \right)}}}	% this is a lighter def of myCl without the scalar product g and with the field
\newcommand{\myClg}[3]{\ensuremath{{{\cal C}\ell} {\left( #3 \right)}}}	% this is a possible def of myCl that uses only the scalar product g

\DeclareMathOperator{\Automorphism}{Aut}	%Automorphism group
\newcommand{\Aut}[1]{\ensuremath{\Automorphism\left(#1\right)}}	%Automorphism group
}

\opt{JMP}{% only for submission to JMP -> double line spacing
}

\def\h_eigen{\eta}
\def\g_eigen{\theta}
\def\mygen{e} % this way the definition of the generators of the algebra can be easily changed
\def\mydual{{*}} % this way the definition of the dual/transpose can be easily changed eg. with {t} or {L} or {*} or {\cal L}
\def\myeee{f} % this way the definition of the all 1's vector e = (1,1, ...,1) can be easily changed

\begin{document}

\opt{x,std,arXiv,JMP,JOPA}{% in all cases - but AACA
\title{{\bf On Spinors Transformations} %\\(temporary title)
	}

\author{\\
	\bf{Marco Budinich}%
%
%\footnote{on leave of absence from: University of Trieste, Trieste, Italy}%
%
\\
%	ICTP and INFN, Trieste, Italy\\
	University of Trieste and INFN, Trieste, Italy\\
	\texttt{mbh@ts.infn.it}\\
%	\texttt{http://www.ts.infn.it/\~{ }mbh/MBHgeneral.html}\\
%
%\\	Very preliminary - restricted circulation (FYEO)
%
%
%	Submitted on March 17, 2016
%	Submitted to: {\em Journal of Mathematical Physics} on March 30, 2016
%		keywords of this submission: spinors, discrete transformations, Clifford algebra
%	Submitted to: {\em Communications in Mathematical Physics} on March 14, 2016.\\
%	Submitted to: {\em Journal of Physics A: Mathematical and Theoretical} \\on March 17, 2016
%	Submitted to: {\em Letters in Mathematical Physics} on July 28, 2011
%	Submitted to: {\em Advances in Applied Clifford Algebras} on September 5, 2011
%	Published in: {\em Advances in Applied Clifford Algebras}, 2015\\
%	{\small DOI:10.1007/s00006-015-0547-8}
%	Resubmitted to: {\em Journal of Mathematical Physics} on June 27, 2016\\
%	{\tiny First submission: March 30, 2016}
%	To appear in: {\em Journal of Mathematical Physics}\\
%	{\tiny Submitted on March 30 and June 27, 2016; accepted July 12, 2016}
	{\em Journal of Mathematical Physics} {\bf 57} (2016) DOI: 10.1063/1.4959531\\
%	\\ May 28, 2014, submitted
	}
\date{ \today }
%\date{June 27, 2016}
%\date{ } % to hide the date this line must be present (believe it or not...)
\maketitle
}

\opt{AACA}{% only for AACA
\title[On Spinors of Zero Nullity]{On Spinors of Zero Nullity}

\author{Marco Budinich}
\address{Dipartimento di Fisica\\
	Università di Trieste \& INFN\\
	Via Valerio 2, I - 34127 Trieste, Italy}
\email{mbh@ts.infn.it}
}

\begin{abstract}
We begin showing that for even dimensional vector spaces $V$ all automorphisms of their Clifford algebras are inner. So all orthogonal transformations of $V$ are restrictions to $V$ of inner automorphisms of the algebra. Thus under orthogonal transformations $P$ and $T$ --- space and time reversal --- all algebra elements, including vectors $v$ and spinors $\varphi$, transform as $v \to x v x^{-1}$ and $\varphi \to x \varphi x^{-1}$ for some algebra element $x$. We show that while under combined $PT$ spinor $\varphi \to x \varphi x^{-1}$ remain in its spinor space, under $P$ or $T$ separately $\varphi$ goes to a \emph{different} spinor space and may have opposite chirality. We conclude with a preliminary characterization of inner automorphisms with respect to their property to change, or not, spinor spaces.
\end{abstract}

%\opt{x,std,arXiv,JMP,JOPA}{% in all cases - but AACA
%\noindent{\bf Keywords:} {Spinors, discrete transformations, Clifford algebra.}
%}

\opt{AACA}{% only for AACA
\maketitle
}

\section{Introduction}
\label{Introduction}
\opt{margin_notes}{\mynote{mbh.note: for paper material see log pp. 640 ff.}}%
In 1913 {\'{E}}lie Cartan introduced spinors \cite{Cartan_1913, Cartan_1937} and, after more than a century, this vein looks inexhaustible.
% mother lode is always rich.
Spinors were later thoroughly investigated by Claude Chevalley \cite{Chevalley_1954} in the mathematical frame of Clifford algebras where they were identified as elements of Minimal Left Ideals (MLI) of the algebra. Many years later Benn and Tucker \cite{Benn_1987} and Porteous \cite{Porteous_1995} wrote books with many of these results easier to be assimilated by physicists.

In this paper we address the transformation properties of spinors under certain inner automorphisms of Clifford algebra exploiting the Extended Fock Basis (EFB) of Clifford algebra \cite{BudinichM_2009, Budinich_2011_EFB} recalled in section~\ref{Clifford_algebra_and_EFB}. As a sample application of this formalism we show how to write vectors as linear superposition of simple spinors (\ref{e_i_in_EFB}), thus supporting the well-known Penrose twistor program \cite{Penrose_1988} that spinor structure is the underlying --- more fundamental --- structure of Minkowski spacetime.

In the subsequent section~\ref{Spinor_spaces} we review some quite general properties of a simple Clifford algebra and in particular the fact that it contains many different MLI, namely many different spinor spaces, that are completely equivalent%
\opt{margin_notes}{\mynote{mbh.ref: Porteous 1995 \cite[p. 133]{Porteous_1995} B\&T \cite[p. ?]{Benn_1987}}}%
{} in the sense that each of them can carry an equivalent representation; moreover the algebra, as a vector space, is the direct sum of these spinor spaces. These properties are known and recently it has been suggested that multiple spinor spaces play a role in physics \cite{Pavsic_2010}.

One of the most important properties of Clifford algebra is that it establishes a deep connection between the orthogonal transformations of vector space $V$ with scalar product $g$ (more precisely: its image in the algebra) and the automorphisms of Clifford algebra \myClg{}{}{g}. In section~\ref{Automorphisms} we show first that if the vector space is even dimensional then all \myClg{}{}{g} automorphisms are \emph{inner} automorphisms and thus that all orthogonal transformations on $V$ lift to inner automorphisms of \myClg{}{}{g}. We then examine in detail the so called discrete orthogonal transformations of $V$, namely $\Identity_V, P, T$ and $PT$ ($V$ identity, space and time reversal and their composition) and we focus on the inner algebra automorphism they induce. This study takes advantage of the properties of the EFB that allow to remain within the algebra without using representations. At the same time we exhibit the elements of the algebra that generate these inner automorphisms. It follows that we can look at $\Identity_V, P, T$ and $PT$ as at restrictions of automorphisms of the entire algebra to $V$, thus unifying the treatment of the discrete transformations of $V$ with those of the continuous ones of the \Pin{g} group.

A similar approach was followed also by Varlamov \cite{Varlamov_2001,Varlamov_2015} to study the hierarchies of \Pin{g} and \OO{g} groups and he succesfully classified the automorphisms of \myClg{}{}{g} showing that the eight double coverings of \OO{g}, the Dabrowski groups \cite{Dabrowski_1988}, correspond to the eight types of real Clifford algebras: the so called ``spinorial clock'' \cite{BudinichP_1988e}.
\opt{margin_notes}{\mynote{mbh.ref: \cite{Varlamov_2001} 667 p. 5}}%

Here we exploit the same unification to investigate a different subject: given an inner automorphism
$$
\alpha: \myClg{}{}{g} \to \myClg{}{}{g}; \;\; \alpha(\mu) = x \mu x^{-1} \qquad x \in \myClg{}{}{g}
$$
it is manifest that all algebra elements must transform accordingly and in particular that the typical physics equations $v \varphi = 0$, where $v \in V$ and $\varphi$ is a spinor, must go to $\alpha(v \varphi) = 0$. We remark that $\varphi$ is \emph{both} a carrier of the regular representation \emph{and} an element of \myClg{}{}{g} so the equation $\alpha(v \varphi) = 0$ is justified. Since the automorphism is inner it follows that both $v$ and spinor $\varphi$ must transform as $\alpha(\varphi) = x \varphi x^{-1}$ thus adding an ``extra'' $x^{-1}$ to the ``traditional rule'' stating that vectors transform as $v \to x v x^{-1}$ while spinors as $\varphi \to x \varphi$.
\opt{margin_notes}{\mynote{mbh.note: this is the only place where we make reference to spinors ``external'' to Clifford algebras}}%
This consequence is unavoidable if we accept that spinors are part of the Clifford algebra and not elements of some ``external'' linear space, a point of view that, even if historical, is rarely taken nowadays.

We examine in detail the spinors transformations $\alpha(\varphi) = x \varphi x^{-1}$ proving that if on one side they can not alter in any way the solutions of $v \varphi = 0$, on the other hand, in some cases, they ``move'' $\varphi$ to a different spinor space, one of the many equivalent ones in \myClg{}{}{g}. In particular we show that while the automorphisms corresponding to $\Identity_V$ and $PT$ do not move spinors, those corresponding to $P$ and $T$ move them, thus populating other spinor spaces.

In section~\ref{Spinors_transformations} we begin the characterization of these automorphisms: those that keep the spinor space constant, like $\Identity_V$ and $PT$ and those that do not, like $P$ and $T$, and we show that the latter transformations can also invert spinor chiralities. This is the first part of the study of these transformations that will be completed in a companion paper where also continuous transformations will be examined.

For the convenience of the reader we tried to make this paper as elementary and self-contained as possible.

\section{Clifford algebra and its 'Extended Fock Basis'}
\label{Clifford_algebra_and_EFB}
We summarize the essential properties of the EFB introduced in 2009 \cite{BudinichM_2009, Budinich_2011_EFB}; we consider Clifford algebras \myClg{}{}{g} \cite{Chevalley_1954, Porteous_1995, BudinichP_1988e} over the fields $\F = \R$ or $\C$ and an \emph{even} dimensional vector space $V$ equipped with a non degenerate scalar product $g$; any base $\mygen_1, \mygen_2, \ldots, \mygen_{n}$ with $n = 2 m$ generates the algebra that results: simple, central and of dimension $2^{n}$.

EFB formalism is fully developed for \emph{neutral} spaces: $V = \C^{2 m}$ or $\R^{m, m}$, spaces for which Witt decomposition is the direct sum of two totally null (isotropic) subspaces of dimension $m$; when we refer to this case we indicate the corresponding Clifford algebra \myCl{m}{m}{}. This choice allows to treat a simpler case, avoiding the many intricacies brought in by other signatures (the extension of the formalism to other cases is under development).
\opt{margin_notes}{\mynote{mbh.ref: exploiting $\C \otimes \myClg{}{}{\R^{k,l}} \myisom \myClg{}{}{\C^{k+l}}$}}%
At the same time this restriction is much milder than it may seem since the following results apply also to the complexification of the Clifford algebras of even dimensional real spaces of \emph{any} signature.

In neutral spaces the $\mygen_i$'s form an orthonormal basis of $V$ with \eg
\opt{margin_notes}{\mynote{mbh.note: remember that $\myCl{2}{2}{} \myisom \myCl{3}{1}{} \myisom \R(4)$ (Dirac algebra ?), while $\myCl{1}{3}{} \myisom \HH(2)$ and for both the even subalgebras $\myisom \C(2)$ ...}}%
$$
2 g(\mygen_i, \mygen_j) = \mygen_i \mygen_j + \mygen_j \mygen_i := \anticomm{\mygen_i}{\mygen_j} := g_{i j} = 2 \delta_{i j} (-1)^{i+1}
$$
while $\anticomm{\mygen^i}{\mygen_j} = 2 \delta^i_j$ and 
\opt{margin_notes}{\mynote{mbh.note: This result is proved exactly in Porteous book 1981 648 Theorem~13.27, p.~ 249.}}%
\opt{margin_notes}{\mynote{mbh.ref: For $\C$ the ``signature'' can be freely chosen, see \eg p. 496' and Math\_643.}}%
\begin{equation}
\label{space_signature}
\left\{ \begin{array}{l l l}
\mygen_{2 i - 1}^2 & = & 1 \\
\mygen_{2 i}^2 & = & -1
\end{array} \right.
\qquad i = 1,\ldots,m \dotinformula
\end{equation}
The Witt, or null, basis of the vector space $V$ is defined, for both fields:
\begin{equation}
\label{formula_Witt_basis}
\left\{ \begin{array}{l l l}
p_{i} & = & \frac{1}{2} \left( \mygen_{2i-1} + \mygen_{2i} \right) \\
q_{i} & = & \frac{1}{2} \left( \mygen_{2i-1} - \mygen_{2i} \right)
\end{array} \right.
\Rightarrow
\left\{\begin{array}{l l l}
\mygen_{2i-1} & = & p_{i} + q_{i} \\
\mygen_{2i} & = & p_{i} - q_{i}
\end{array} \right.
\quad i = 1,2, \ldots, m
\end{equation}
that, with $\mygen_{i} \mygen_{j} = - \mygen_{j} \mygen_{i}$, gives
\begin{equation}
\label{formula_Witt_basis_properties}
\anticomm{p_{i}}{p_{j}} = \anticomm{q_{i}}{q_{j}} = 0
\qquad
\anticomm{p_{i}}{q_{j}} = \delta_{i j}
\end{equation}
showing that all $p_i, q_i$ are mutually orthogonal, also to themselves, that implies $p_i^2 = q_i^2 = 0$, at the origin of the name ``null'' given to these vectors.

Following Chevalley we define spinors as elements of a MLI $S$; \emph{simple} (pure) spinors are those elements of $S$ that are annihilated by a null subspace of $V$ of maximal dimension $m$.

\bigskip

The EFB of \myCl{m}{m}{} is given by the $2^{2 m}$ different sequences
\begin{equation}
\label{EFB_def}
\psi_1 \psi_2 \cdots \psi_m := \Psi \qquad \psi_i \in \{ q_i p_i, p_i q_i, p_i, q_i \} \qquad i = 1,\ldots,m
\end{equation}
in which each $\psi_i$ can take four different values and we reserve $\Psi$ for EFB elements and $\psi_i$ for its components. The main characteristics of EFB is that all its $2^{2 m}$ elements $\Psi$ are simple spinors \cite{BudinichP_1989}.

The EFB essentially extends to the entire algebra the Fock basis of its spinor spaces and, making explicit the relation $\myCl{m}{m}{} \myisom \overset{m}{\otimes} \myCl{1}{1}{}$, allow to trace back in \myCl{1}{1}{} many properties of \myCl{m}{m}{}. We stress that this constitutes a base of the algebra itself and not of its representations and the matrix formalism, with row and column indices, emerges right from the algebra.

\subsection{$h$ and $g$ signatures}
\label{hg_signatures}
We start observing that $\mygen_{2 i - 1} \mygen_{2 i} = q_i p_i - p_i q_i := \comm{q_i}{p_i}$ and that for $i \ne j$ $\comm{q_i}{p_i} \psi_j = \psi_j \comm{q_i}{p_i}$. With (\ref{formula_Witt_basis_properties}) and (\ref{EFB_def}) it is easy to calculate
\begin{equation}
\label{commutator_property}
\comm{q_i}{p_i} \psi_i = h_i \psi_i \qquad h_i = \left\{
\begin{array}{l l}
+1 & \quad \mbox{iff $\psi_i = q_i p_i \; \mbox{or} \; q_i$} \\
-1 & \quad \mbox{iff $\psi_i = p_i q_i \; \mbox{or} \; p_i$}
\end{array} \right.
\end{equation}
and the value of $h_i$ depends on the first null vector appearing in $\psi_i$. We have thus proved that $\comm{q_i}{p_i} \Psi = h_i \Psi$ and thus each EFB element $\Psi$ defines a vector $h = (h_1, h_2, \ldots, h_m) \in \{ \pm 1 \}^m$ that we name ``$h$ signature''. In EFB the \myCl{m}{m}{} identity $\Identity$ and the volume element $\omega$ (scalar and pseudoscalar) have similar expressions:
\begin{equation}
\label{identity_omega}
\begin{array}{l l l}
\Identity & := & \anticomm{q_1}{p_1} \anticomm{q_2}{p_2} \cdots \anticomm{q_m}{p_m} \\
\omega & := & \mygen_1 \mygen_2 \cdots \mygen_{2 m} = \comm{q_1}{p_1} \comm{q_2}{p_2} \cdots \comm{q_m}{p_m}
\end{array}
\end{equation}
with which and defining a function $\epsilon: \{ \pm 1 \}^m \to \{ \pm 1 \}, \epsilon(h) = \prod_{i = 1}^m h_i$
\begin{equation}
\label{Weyl_eigenvector}
\omega \Psi = \h_eigen \; \Psi \qquad \h_eigen := \epsilon(h) = \pm 1 \dotinformula
\end{equation}
Each EFB element $\Psi$ has thus an eigenvalue $\h_eigen$: the {\em chirality}. Similarly the ``$g$ signature'' of an EFB element is the vector $g = (g_1, g_2, \ldots, g_m) \in \{ \pm 1 \}^m$ (not to be confused with scalar product $g$) where $g_i$ is the parity of $\psi_i$ under the main algebra automorphism $\alpha(\mygen_i) = - \mygen_i$ (\ref{main_automorphism_def}). With this definition and with (\ref{commutator_property}) we easily obtain
\begin{equation}
\label{commutator_left_property}
\psi_i \comm{q_i}{p_i} = g_i \comm{q_i}{p_i} \psi_i = h_i g_i \psi_i
\end{equation}
and thus
\begin{equation}
\label{EFB_are_left_Weyl}
\Psi \; \omega = \h_eigen \g_eigen \; \Psi \qquad \h_eigen \g_eigen = \pm 1 \qquad \g_eigen := \epsilon(g)
\end{equation}
\opt{margin_notes}{\mynote{mbh.note: perhaps here one could calculate $\Psi \; \omega^{-1}$, see comments at (\ref{omega_transforms}) and log p. 659}}%
where the eigenvalue $\h_eigen \g_eigen$ is the product of chirality times $\g_eigen$, the global parity of the EFB element $\Psi$ under the main algebra automorphism. We can resume saying that all EFB elements are not only Weyl eigenvectors, \ie right eigenvectors of $\omega$ (\ref{Weyl_eigenvector}), but also its left eigenvectors (\ref{EFB_are_left_Weyl}) with respective eigenvalues $\h_eigen$ and $\h_eigen \g_eigen$, a property we will use in what follows.

\subsection{EFB formalism}
\label{EFB_formalism}
$h$ and $g$ signatures play a crucial role in this description of \myCl{m}{m}{}: first of all we notice that any EFB element $\Psi = \psi_1 \psi_2 \cdots \psi_m$ is uniquely identified by its $h$ and $g$ signatures: $h_i$ determines the first null vector ($q_i$ or $p_i$) appearing in $\psi_i$ and $g_i$ determines if $\psi_i$ is even or odd, see (\ref{EFB_def}).

It can be shown \cite{Budinich_2011_EFB} that \myCl{m}{m}{}, as a vector space, is the direct sum of its $2^m$ subspaces of:
\begin{itemize}
\item different $h$ signatures or:
\item different $g$ signatures or:
\item different $h \circ g$ signatures, where $h \circ g \in \{ \pm 1 \}^m$ is the Hadamard (entrywise) product of $h$ and $g$ signature vectors; $h \circ g = (h_1 g_1, \ldots, h_m g_m)$.
\end{itemize}

We can thus uniquely identify each of the $2^{2 m}$ EFB elements with any two of these three ``indices''. For reasons that will be clear in a moment we choose the $h$ and the $h \circ g$ signatures \ie
\begin{equation}
\label{EFB_index_def}
\Psi_{a b} \left\{
\begin{array}{l l}
a \in \{ \pm 1 \}^m \quad \mbox{is the} \quad h \mbox{ signature} \\
b \in \{ \pm 1 \}^m \quad \mbox{is the} \quad h \circ g \mbox{ signature}
\end{array} \right.
\end{equation}
so that the generic element of $\mu \in \myCl{m}{m}{}$ can be written as $\mu = \sum_{a b} \xi_{a b} \Psi_{a b}$ with $\xi_{a b} \in \F$. With this choice of the indices it can be proved \cite{Budinich_2011_EFB} that:
\begin{equation}
\label{EFB_products}
\Psi_{a b} \Psi_{c d} = s(a,b,d) \, \delta_{b c} \Psi_{a d} \qquad s(a,b,d) = \pm 1
\end{equation}
where $\delta_{b c}$ is $1$ if and only if the two signatures $b$ and $c$ are equal and the sign $s(a,b,d)$, slightly tedious to calculate, depends on the indices; in \cite{Budinich_2011_EFB} it is shown how it can be calculated recursively. This formula explains the choice of $h \circ g$ signature since it is now clear that different $h \circ g$ signatures identify different MLI and thus different spinor spaces, denoted $S_{hg}$ for short. We can thus calculate the most general Clifford product
\begin{eqnarray*}
\mu \nu & = & \left(\sum_{a b} \xi_{a b} \Psi_{a b} \right) \left(\sum_{c d} \zeta_{c d} \Psi_{c d} \right) = \sum_{a b c d} \xi_{a b} \zeta_{c d} \Psi_{a b} \Psi_{c d} = \\
& = & \sum_{a d} \Psi_{a d} \sum_{b} s(a,b,d) \xi_{a b} \zeta_{b d} := \sum_{a d} \rho_{a d} \Psi_{a d}
\end{eqnarray*}
having defined $\rho_{a d} = \sum_{b} s(a,b,d) \xi_{a b} \zeta_{b d}$.

So EFB elements naturally display a matrix structure, mirrored in the isomorphic full matrix algebra $\F( 2^m )$, where $a$ and $b$ are respectively the row and column indices of $\Psi_{a b}$ when interpreted as binary numbers substituting: $1 \to 0$ and $ -1 \to 1$. Let
$$
\myeee := (-1,-1,-1,\ldots,-1) \in \{ \pm 1 \}^m
$$
then, with the proposed substitutions, $- \myeee$ gives the binary expression of $0$ and $\myeee$ that of $2^m - 1$, see \cite{Budinich_2011_EFB}; moreover by (\ref{Weyl_eigenvector}), ({\ref{EFB_are_left_Weyl}) and (\ref{EFB_index_def})
\opt{margin_notes}{\mynote{mbh.note: obviously $\h_eigen \g_eigen = \epsilon(b)$ is constant in any spinor space}}%
\begin{equation}
\label{chirality_parity}
\begin{array}{l l l}
\h_eigen(\Psi_{a b}) & = & \epsilon(a) \\
\g_eigen(\Psi_{a b}) & = & \epsilon(a) \epsilon(b) \dotinformula
\end{array}
\end{equation}

As an example we give the EFB for $\Cl(2,2) \myisom \myCl{3}{1}{} \myisom \R(4)$ with $h$ (rows) and $h \circ g$ (columns) signatures (taken from \cite{Budinich_2011_EFB}):
\begin{equation}
\label{EFB_Cl_2_2_example}
\bordermatrix{& ++ & +- & -+ & -- \cr
++ & q_1 p_1 \, q_2 p_2 & q_1 p_1 \, q_2 & q_1 \, q_2 p_2 & q_1 \, q_2 \cr
+- & q_1 p_1 \, p_2 & q_1 p_1 \, p_2 q_2 & - q_1 \, p_2 & - q_1 \, p_2 q_2 \cr
-+ & p_1 \, q_2 p_2 & p_1 \, q_2 & p_1 q_1 \, q_2 p_2 & p_1 q_1 \, q_2 \cr
-- & - p_1 \, p_2 & - p_1 \, p_2 q_2 & p_1 q_1 \, p_2 & p_1 q_1 \, p_2 q_2 \cr }
\end{equation}
where the signs of matrix elements come from (\ref{EFB_products}). With (\ref{formula_Witt_basis}) and (\ref{EFB_def}) we can write the standard $\mygen_i$ base in EFB as a sum of $2^{m} = 4$ EFB terms
\opt{margin_notes}{\mynote{mbh.note: signs ok with (\ref{dual_transforms}): only $g$ odd-odd change sign in transposition}}%
\begin{eqnarray*}
\mygen_1 = (p_1 + q_1) = (p_1 + q_1) \anticomm{p_{2}}{q_{2}} \\
\mygen_2 = (p_1 - q_1) = (p_1 - q_1) \anticomm{p_{2}}{q_{2}} \\
\mygen_3 = (p_2 + q_2) = \anticomm{p_{1}}{q_{1}} (p_2 + q_2) \\
\mygen_4 = (p_2 - q_2) = \anticomm{p_{1}}{q_{1}} (p_2 - q_2)
\end{eqnarray*}
and with (\ref{EFB_Cl_2_2_example}) we can write their matrix forms
\begin{eqnarray*}
\mygen_1 = \left(\begin{array}{r r r r} 0 & 0 & 1 & 0 \\ 0 & 0 & 0 & -1 \\ 1 & 0 & 0 & 0 \\ 0 & -1 & 0 & 0 \end{array}\right)
\mygen_2 = \left(\begin{array}{r r r r} 0 & 0 & -1 & 0 \\ 0 & 0 & 0 & 1 \\ 1 & 0 & 0 & 0 \\ 0 & -1 & 0 & 0 \end{array}\right) \\
\mygen_3 = \left(\begin{array}{r r r r} 0 & 1 & 0 & 0 \\ 1 & 0 & 0 & 0 \\ 0 & 0 & 0 & 1 \\ 0 & 0 & 1 & 0 \end{array}\right)
\mygen_4 = \left(\begin{array}{r r r r} 0 & -1 & 0 & 0 \\ 1 & 0 & 0 & 0 \\ 0 & 0 & 0 & -1 \\ 0 & 0 & 1 & 0 \end{array}\right)
\end{eqnarray*}
\opt{margin_notes}{\mynote{mbh.note: after a quick look they do not resemble neither Dirac, nor Weyl, nor Majorana $\gamma$ matrices}}%
where we interpret the nonzero terms of these matrices as the simple spinors building up vectors $\mygen_i$'s; moreover it's simple to identify the two null vectors $p_i$ and $q_i$ in each of them. We can exploit these expressions further to prove some fairly general properties of $\mygen_i$'s for any $m$; writing
\begin{equation}
\label{e_i_in_EFB}
\mygen_i = \left(p_i + (-1)^{i + 1} q_i \right) \prod_{\stackrel{j = 1}{j \ne i}}^m \anticomm{p_{j}}{q_{j}}
\end{equation}
we notice that it expands in a sum of exactly $2^m$ simple spinors, all with identical $g$ signature $g = (1, \ldots, 1, -1, 1, \ldots, 1)$ with the only $-1$ at the $i$-th place. It is clear that EFB terms of the sum cover all $2^m$ possible $h$ signatures and all possible $h \circ g$ signatures, each ``column'' and each ``row'' being filled exactly once in a pattern similar to that of a permutation matrix.%
\opt{margin_notes}{\mynote{mbh.ref: see log. p. 647}}%

To show the power of this formulation we prove that \eg $\mytrace{\mygen_i} = 0$: we see immediately that all diagonal terms, those with identical $h$ and $h \circ g$ signatures, are forbidden since $g \ne -\myeee$. We could continue proving, within the algebra, other familiar properties of gamma matrices.

As a side remark we observe that this formulation provides the faster algorithm for actual Clifford product evaluations \cite{BudinichM_2009} resulting a factor $2^m$ faster than algorithms based on gamma matrices.

\opt{margin_notes}{\mynote{mbh.note: can we continue, as in Porteous book 1981 648, building base-free matrices, namely interpreting simple spinors as endomorphisms of $S$ ?}}%
We have shown that for neutral spaces matrix multiplication rules are integral part of Clifford algebra without the need to resort to representations.

\section{Multiple spinor spaces}
\label{Spinor_spaces}
We already mentioned that \myCl{m}{m}{}, as a vector space, is the direct sum of subspaces of different $h \circ g$ signatures \cite{Budinich_2011_EFB}. Given the Clifford product properties (\ref{EFB_products}) these subspaces are also MLI of \myCl{m}{m}{} and thus coincide with $2^m$ different spinor spaces $S_{hg}$ that in turn correspond to different columns of the isomorphic matrix algebra $\F( 2^m )$. To establish a further link between EFB and the familiar definition of MLI in \myCl{m}{m}{} \cite{Benn_1987} we remark that in EFB the $2^m$ elements with identical $h$ and $h \circ g$ signatures, namely $\Psi_{a a}$, are primitive idempotents and the MLI $S_a$ can thus be written as $S_a = \myCl{m}{m}{} \Psi_{a a}$.%
\opt{margin_notes}{\mynote{mbh.ref: see \eg 666 \cite{Varlamov_2015} p. 13 and Lounesto's book 438 p. 226}}%

Each of the $2^m$ spinor spaces supports a regular, faithful and irreducible representation of \myCl{m}{m}{} and since the algebra is simple there exist isomorphisms intertwining the representations. This has been known for a long time but recently mirror particles \cite{Pavsic_2010} have been proposed as a possible realization of multiple spinor spaces.
\opt{margin_notes}{\mynote{mbh.ref: see references in box at p. 2}}%
Here we show how, under certain transformations, \eg $P$ and $T$, a spinor moves to a different spinor space.

We choose a particular spinor space, \eg $h \circ g = \myeee$, the rightmost column in example (\ref{EFB_Cl_2_2_example}), so that when speaking of a generic spinor we will refer to spinor space $S_{\myeee}$ (used to build the Fock basis in \cite{BudinichP_1989}). Its generic element $\varphi \in S_{\myeee}$ can thus be expanded in the Fock basis
\begin{equation}
\label{Fock_basis_expansion}
\varphi = \sum_{a} \xi_{a \myeee} \Psi_{a \myeee}
\end{equation}
and, since the second index $\myeee$ is constant, in principle it could be omitted. 

Let this $\varphi \in S_{\myeee}$ be a solution of the Weyl equation $v \varphi = 0$ where $v \in V$; we remark that the equation is solved also by all $\varphi' = \sum_{a} \xi_{a} \Psi_{a \myeee'} \in S_{\myeee'}$ for any $\myeee'$, this being a simple consequence of (\ref{EFB_products}); we will return to this point in paragraph~\ref{Spinors_transformations}.

\subsection{Representations of Clifford algebra \myClg{}{}{g}}
\label{Cl_representations}
\opt{margin_notes}{\mynote{mbh.note: copied heavily from Trautman's 610 \cite{Trautman_2006} p. 7}}%
We resume some quite general properties we need in the sequel: let $\gamma : \myClg{}{}{g} \to \End S$ be a faithful irreducible representation of \myClg{}{}{g} and let $\beta$ be the so called main antiautomorphism
\begin{equation}
\label{main_antiautomorphism_def}
\left\{ \begin{array}{l l l l}
\beta(\mu \nu) & = & \nu \mu & \quad \forall \: \mu, \nu \in \myClg{}{}{g} \\
\beta(v) & = & v & \quad \forall \: v \in V \\
\beta(\Identity) & = & \Identity
\end{array} \right.
\end{equation}
that reverses the order of multiplication and that is involutive. With $\beta$ it is possible to define the contragredient representation in $S^{\mydual}$, the dual of $S$, $\check{\gamma} : \myClg{}{}{g} \to \End S^{\mydual}$ given by $\check{\gamma}(\mu) = \gamma\left(\beta(\mu)\right)^{\mydual}$ and, since in our case $V$ is even dimensional, \myClg{}{}{g} is simple and central and thus there exists an isomorphism $B : S \to S^{\mydual}$ intertwining the two representations: $\check{\gamma} B = B \gamma$ which is either symmetric $B = B^{\mydual}$ or antisymmetric $B = - B^{\mydual}$ \cite{BudinichP_1988e, Trautman_2006, Case_1955}%
\opt{margin_notes}{\mynote{mbh.note: if algebra not central this holds only for $\C$; see Spinorial Chessboard p. 43 and log p. 254}}%
{} and that also defines on $S$ the structure of an inner product space ($\langle \cdot , \cdot \rangle$ represents the bilinear product or contraction)
$$
S \times S \to \F \qquad B(\varphi, \phi) := \langle B \varphi, \phi \rangle \in \F \dotinformula
$$
This structure extends to $\End S$: there is a symmetric isomorphism $B \otimes B^{-1} : \End S \to (\End S)^{\mydual} = \End S^{\mydual}$ given, for every $\gamma \in \End S$, by $(B \otimes B^{-1})(\gamma) = B \gamma B^{-1}$.
\opt{margin_notes}{\mynote{mbh.note: via canonical isomorphism $S \otimes S^{\mydual} \myisom \End_\F S$ see p. 387' that can also be used to show that $S \otimes S^{\mydual} \myisom \End_\F S^{\mydual}$}}%
These results are fully general and hold thus also when $\gamma$ is the regular representation $\gamma(\mu) = \mu$, $\End S = \myClg{}{}{g}$ and $S$ is one of its MLI.

\section{Automorphisms of Clifford algebra \myClg{}{}{g}}
\label{Automorphisms}

We begin with a general proposition and thus in this section there are no restrictions on the dimensions of the vector space $V$.
\begin{MS_Proposition}
\label{Cl_automorphisms_thm}
For a Clifford algebra over fields $\R$ and $\C$ all its automorphisms are inner if and only if the dimension of the vector space is even.
\end{MS_Proposition}
\begin{proof}
For any non degenerate, even dimensional, vector space $V$ its Clifford algebra is central and simple \cite{BudinichP_1988e} and, by Skolem -- Noether theorem, all its automorphisms are inner. To prove the converse we take an odd dimensional vector space and we examine the so called main automorphism of its Clifford algebra, the automorphism that reverses all vectors (\ref{main_automorphism_def}). In this case the volume element $\omega$ (\ref{identity_omega}) is formed by an odd number of vectors and thus the main automorphism sends $\omega \to -\omega$. But in this case $\omega$ belongs also to the center of the algebra and thus for any inner automorphism $x \omega x^{-1} = \omega$, thus the main automorphism is not inner.
\end{proof}

A simple example is $\myClf{0}{1}{\R} \myisom \C$ where the main automorphism coincides with complex conjugation and is not inner.
%In general the Clifford algebra of an odd dimensional vector space has the form $\myClg{}{}{g} \oplus \myClg{}{}{g}$ and has no inner elements giving the main automorphism, nevertheless the main automorphism is obtained by the ``swap'' automorphism $\left(\begin{array}{c c} 0 & \Identity \\ \Identity & 0 \end{array}\right)$ that \emph{is not} an element of the algebra but \emph{is} an element of the matrix representations of the algebra.

A corollary that follows from the universality of Clifford algebra is that all orthogonal transformations on an even dimensional $V$ lift to inner automorphisms of \myClg{}{}{g}. This corollary gives a simpler proof, but only for even dimensional spaces, of the result quoted in \cite{Varlamov_2001} that all ``fundamental automorphisms'', even discrete ones like $P$ and $T$, are inner automorphisms.

So in even dimensional spaces
\opt{margin_notes}{\mynote{mbh.note: the exp map maps \myClg{}{}{g} in $C_g^{\star}$; moreover $V + \comm{u}{v}$ identifies with the Lie algebra of $C_g^{\star}$; see B\&T \cite[p.50]{Benn_1987}}}%
\begin{equation}
\label{Cg_star_def}
\Aut{\myClg{}{}{g}} = \{x \in \myClg{}{}{g} : \exists x^{-1} \} := C_g^{\star}
\end{equation}
\opt{margin_notes}{\mynote{mbh.note: Here $\anticomm{u}{v}$ is preserved by $x v x^{-1}$ but it may $\notin V$. See pp. 356.1, 566. About conflicting definition of the Clifford Lipschitz group see B\&T \cite[p.45]{Benn_1987}}}%
and the Clifford Lipschitz group is its subgroup that stabilizes vectors that in turn, when restricted on vector space, is the orthogonal group \OO{g}.

\subsection{Fundamental automorphisms of \myClg{}{}{g}}
\label{Fundamental_automorphisms}
In general in \myClg{}{}{g} there are four automorphisms corresponding to the two involutions and to the two antinvolutions induced by the orthogonal transformations $\Identity_V$ and $-\Identity_V$ of vector space $V$ \cite[Theorem~15.32]{Porteous_1995}. They are called fundamental or discrete automorphisms and form a finite group, isomorphic to the Klein four group $\Z_2 \otimes \Z_2$ \cite{Varlamov_2001}. We review them briefly to show how they appear in EFB formalism and to exhibit the elements of $C_g^{\star}$ realizing the inner automorphisms in even dimensional vector spaces.

From now on we restrict again to even dimensional, neutral, spaces to fully exploit EFB properties.

\subsection{Identity automorphism of \myClg{}{}{g}}
\label{Identity_automorphism}
The just quoted theorem proves also that $\Identity_V$ induces the algebra identity automorphism $\Identity$, its internal element in \myCl{m}{m}{} is given in (\ref{identity_omega}).

\subsection{Main automorphism of \myClg{}{}{g}}
\label{Main_automorphism}
\opt{margin_notes}{\mynote{mbh.ref: Porteous book 1981 648 Theorem~13.31, p.~ 252}}%
The main automorphism $\alpha$ of \myClg{}{}{g} (main involution in \cite{Porteous_1995}) is induced by $V$ orthogonal transformation $-\Identity_V$, namely
\begin{equation}
\label{main_automorphism_def}
\alpha(v) = - v \quad \forall \: v \in V
\end{equation}
it is involutive and defines the basic $\Z_2$ grading of \myClg{}{}{g}.

It's easy to see that given the volume element $\omega = \mygen_1 \cdots \mygen_n$ we obtain $\omega \mygen_i = (-1)^{(n - 1)} \mygen_i \omega$ and so, for even dimensional spaces, we have $\omega v \omega^{-1} = - v$ for any $v \in V$, where $\omega^{-1} = \omega^3 = \pm \omega$ and thus in this case the main automorphism on the entire algebra may be written as:
$$
\alpha(\mu) = \omega \mu \omega^{-1} \dotinformula
$$
For the EFB expansion $\mu = \sum_{a b} \xi_{a b} \Psi_{a b}$ we find with (\ref{Weyl_eigenvector}), (\ref{EFB_are_left_Weyl}) and (\ref{chirality_parity})
\opt{margin_notes}{\mynote{mbh.note: this formula differs from that deducible by (\ref{Weyl_eigenvector}) and (\ref{EFB_are_left_Weyl}) that have just $\omega$ at the right. This is no contradiction since here we are (again) in neutral space hypothesis and in this case $\omega^{-1} = \omega$; anyhow this version is more ``general''; log p. 659}}%
\begin{equation}
\label{omega_transforms}
\alpha(\Psi_{a b}) = \omega \Psi_{a b} \omega^{-1} = \g_eigen_{a b} \Psi_{a b} = \epsilon(a) \epsilon(b) \Psi_{a b}
\end{equation}
where $\g_eigen_{a b} = \pm 1$ is the global parity of the EFB element $\Psi_{a b}$ defined in (\ref{EFB_are_left_Weyl}).

We can double check this formula verifying that vectors, when written in EFB formalism, satisfy (\ref{main_automorphism_def}): let us take \eg the vector $\mygen_i$, by (\ref{e_i_in_EFB}) it can be written in EFB as a sum of $2^{m}$ EFB terms
$$
\omega \mygen_i \omega^{-1} = \omega \left(p_i + (-1)^{i + 1} q_i \right) \prod_{\stackrel{j = 1}{j \ne i}}^m \anticomm{p_{j}}{q_{j}} \omega^{-1}
$$
and since $\omega^{-1} = \pm \omega$, for any $j$ $\anticomm{q_j}{p_j} \omega^{-1} = \omega^{-1} \anticomm{q_j}{p_j}$
$$
\omega \mygen_i \omega^{-1} = \omega \left(p_i + (-1)^{i + 1} q_i \right) \omega^{-1} \prod_{\stackrel{j = 1}{j \ne i}}^m \anticomm{p_{j}}{q_{j}} = - \mygen_i
$$
that generalizes at once to any vector.

\subsection{Reversion automorphism of \myClg{}{}{g}}
\label{Main_antiautomorphism}
The main antiautomorphism $\beta$ (\ref{main_antiautomorphism_def}) is the antiautomorphism induced by the orthogonal transformation $\Identity_V$ of $V$ (reversion antiautomorphism in \cite{Porteous_1995}) and gives an automorphism when ``transposed to the dual space''. If $S_{\myeee}$ is a MLI of \myClg{}{}{g}, the space of spinors, and $\gamma$ the regular representation $\gamma(\mu) = \mu$, then $\check{\gamma}(\mu) = \left( \beta(\mu) \right)^{\mydual}$ is the contragredient representation that defines also the reversion automorphism; with (\ref{main_antiautomorphism_def}) we get its main property:
$$
\check{\gamma}(\mygen_{i_1} \cdots \mygen_{i_k}) = \mygen_{i_1}^{\mydual} \cdots \mygen_{i_k}^{\mydual} \dotinformula
$$
Since it is an automorphism it must be inner, thus there exists $\tau \in \myClg{}{}{g}$ such that $\check{\gamma}(\mu) = \tau \mu \tau^{-1}$ and $\tau$ is fully defined by its action on the generators $\check{\gamma}(\mygen_{i}) = \mygen_{i}^{\mydual} = \tau \mygen_{i} \tau^{-1}$
\opt{margin_notes}{\mynote{mbh.note: see Porteous book 1981 648 Proposition~6.38, log. p. 569.}}%
and since $\anticomm{\mygen_{i}^{\mydual}}{\mygen_j} = g^{i}_{j} = 2 \delta^{i}_{j}$ it follows that
$$
\tau \mygen_{i} \tau^{-1} = \mygen_{i}^{\mydual} = \mygen_{i}^{-1} = \mygen_{i}^{3} \dotinformula
$$
\opt{margin_notes}{\mynote{mbh.note: I see no reasons why this shouldn't hold also in \C}}%
With this result and remembering (\ref{space_signature}) and (\ref{formula_Witt_basis}) it is simple to get the explicit form of $\tau$ that depends on the parity of $m$
\begin{eqnarray*}
\tau & = & \left\{
\begin{array}{l l}
\mygen_{2} \mygen_{4} \cdots \mygen_{2 m} & \quad \mbox{for} \; m \; \mbox{even} \\
\mygen_{1} \mygen_{3} \cdots \mygen_{2 m - 1} & \quad \mbox{for} \; m \; \mbox{odd}
\end{array} \right. \\
& = & \left(p_1 + s \, q_1\right) \left(p_2 + s \, q_2\right) \cdots \left(p_m + s \, q_m\right) \qquad s = (-1)^{m + 1} \dotinformula
\end{eqnarray*}

To evaluate the reversion automorphism on EFB elements we easily get
\begin{equation}
\label{main_antiautomorphism_EFB_def}
\check{\gamma}(\Psi_{a b}) = \beta(\Psi_{a b})^{\mydual} = \beta(\psi_1 \psi_2 \cdots \psi_m)^{\mydual} = \beta({\psi_1})^{\mydual} \beta({\psi_2})^{\mydual} \cdots \beta({\psi_m})^{\mydual} \dotinformula
\end{equation}
By (\ref{main_antiautomorphism_def}) $\beta(p_i) = p_i$, $\beta(q_i) = q_i$ so that $\beta(p_i q_i) = q_i p_i$ and $\beta(q_i p_i) = p_i q_i$. Since $\mygen_{i}^{\mydual} = \mygen_{i}^{-1}$ by (\ref{formula_Witt_basis}) we obtain that $(p_i)^{\mydual} = q_i$, $(p_i q_i)^{\mydual} = p_i q_i$ and $(q_i p_i)^{\mydual} = q_i p_i$. We can resume saying that the $g_{i}$ and $h_{i}$ signatures of $\beta({\psi_i})^{\mydual}$ are respectively equal to $g_i$ and $- h_i$ of that of $\psi_i$ so that the effect of reversion automorphism is to change sign to both $h$ and $h \circ g$ signatures.
\opt{margin_notes}{\mynote{mbh.ref: see pp. 385 and 392}}%
We can thus conclude that for the reversion automorphism we have
\begin{equation}
\label{tau_transforms}
\beta(\Psi_{a b})^{\mydual} = \tau \Psi_{a b} \tau^{-1} = \Psi_{-a -b} % \qquad s'(a,b) = \pm 1
\end{equation}
and we remark that while $\Psi_{a b}$ belongs to spinor space $S_b$, $\beta(\Psi_{a b})^{\mydual}$ belongs to $S_{-b}$, always a \emph{different} spinor space, the main result of this paper.

For completeness we report the results of similar exercises:
\opt{margin_notes}{\mynote{mbh.ref: $s'(a,b)$ calculation done at p. 385.2}}%
\begin{equation}
\label{antiautomorphism_transforms}
\beta(\Psi_{a b}) = s'(a,b) \Psi_{-b -a} \qquad s'(a,b) = \pm 1
\end{equation}
where the sign $s'(a,b)$, straightforward, if slightly tedious to calculate, depends on the indices; it is easy to double check that it satisfies the properties of the main antiautomorphism (\ref{main_antiautomorphism_def}). We also obtain
\begin{equation}
\label{dual_transforms}
\Psi_{a b}^{\mydual} = s'(a,b) \Psi_{b a} \qquad s'(a,b) = \pm 1
\end{equation}
that could also be deduced directly from the natural matrix structure of the EFB with (\ref{EFB_products}); combining both these formulas we reobtain (\ref{tau_transforms}). Since both (\ref{antiautomorphism_transforms}) and (\ref{dual_transforms}) are involutive we have
$$
s'(a,b) = s'(b,a) = s'(-b,-a) = s'(-a,-b) \dotinformula
$$

\subsection{Conjugation automorphism of \myClg{}{}{g}}
\label{conjugation_automorphism}
The composition of the main (\ref{omega_transforms}) and reversion automorphisms (\ref{tau_transforms}) is called conjugation and results in
\opt{margin_notes}{\mynote{mbh.note: see Porteous book 1981 648 Proposition~13.32 and log. p. 601, NB that $\omega$ and $\tau$ do not commute in general; see paper on binary \myClg{}{}{g}}}%
\begin{equation}
\label{conjugation_transforms}
\alpha \left( \beta(\Psi_{a b})^{\mydual} \right) = \omega \tau \Psi_{a b} (\omega \tau)^{-1} = \g_eigen_{a b} \Psi_{-a -b}
\end{equation}
since $\g_eigen_{-a -b} = \g_eigen_{a b}$ given that also this automorphism is involutive; clearly $\alpha \left( \beta(\mygen_{i})^{\mydual} \right) = - \mygen_{i}^{-1}$.

\subsection{A simple example in \myCl{1}{1}{}}
\label{2_dim_example}
We conclude with a simple example in $\myCl{1}{1}{} \myisom \R(2)$ where the EFB is formed by 4 elements: $\{ qp_{+ +}, pq_{- -}, p_{- +}, q_{+ -} \}$ with the subscripts indicating respectively $h$ and $h \circ g$ signatures; its EFB matrix is
\begin{equation*}
\bordermatrix{& + & - \cr
+ & q p & q \cr
- & p & p q \cr }
\end{equation*}
and the generic element $\mu \in \myCl{1}{1}{}$ can be written as
$$
\mu = \xi_{+ +} qp_{+ +} + \xi_{- -} pq_{- -} + \xi_{- +} p_{- +} + \xi_{+ -} q_{+ -} \qquad \xi \in \F
$$
and the application of the three inner automorphisms gives
$$
\begin{array}{l l l}
\omega \mu \omega^{-1} & = & \xi_{+ +} qp_{+ +} + \xi_{- -} pq_{- -} - \xi_{- +} p_{- +} - \xi_{+ -} q_{+ -} \\
\tau \mu \tau^{-1} & = & \xi_{- -} qp_{+ +} + \xi_{+ +} pq_{- -} + \xi_{+ -} p_{- +} + \xi_{- +} q_{+ -} \\
\omega \tau \mu (\omega \tau)^{-1} & = & \xi_{- -} qp_{+ +} + \xi_{+ +} pq_{- -} - \xi_{+ -} p_{- +} - \xi_{- +} q_{+ -}
\end{array}
$$
and $\omega = \mygen_1 \mygen_2 = \comm{q}{p}$, $\tau = \mygen_1 = p + q$, $\omega \tau = - \mygen_2 = q - p$ and $\Identity = \mygen_1^2 = \anticomm{q}{p}$. For comparison, the same automorphisms applied to the standard $\mygen_i$ formulation gives the ordinary results
$$
\begin{array}{l l l}
\mu & = & \xi_{0} \Identity + \xi_{1} \mygen_1 + \xi_{2} \mygen_2 + \xi_{1 2} \mygen_1 \mygen_2 \qquad \xi \in \F \\
\omega \mu \omega^{-1} & = &\xi_{0} \Identity - \xi_{1} \mygen_1 - \xi_{2} \mygen_2 + \xi_{1 2} \mygen_1 \mygen_2 \\
\tau \mu \tau^{-1} & = & \xi_{0} \Identity + \xi_{1} \mygen_1 - \xi_{2} \mygen_2 - \xi_{1 2} \mygen_1 \mygen_2 \\
\omega \tau \mu (\omega \tau)^{-1} & = & \xi_{0} \Identity - \xi_{1} \mygen_1 + \xi_{2} \mygen_2 - \xi_{1 2} \mygen_1 \mygen_2 \dotinformula
\end{array}
$$

\section{Spinors transformations}
\label{Spinors_transformations}
\opt{margin_notes}{\mynote{mbh.note: beware here c.c. vs algebra over \C {} vs $\C \otimes \myClg{}{}{g}$}}%
We begin by observing that in the complex case, and also in complex representations of the real case, also complex conjugation is responsible for an automorphism and things get more complicated: the finite group of fundamental automorphism doubles its size and has been examined in detail in \cite{Varlamov_2015}. We leave aside for the moment complex conjugation and examine what's going on in our simpler case since it is enough for our purpose of introducing general properties of spinor transformations.
\opt{margin_notes}{\mynote{mbh.ref: see ref 666 \cite{Varlamov_2015} p. 6}}%

The inner automorphisms of section~\ref{Automorphisms} are fully general and their restrictions to $V$ correspond to the $V$ transformations: $\Identity_V$, $P$, $T$ and $PT$. We already saw that the restriction of the algebra identity to $V$ is $\Identity_V$ while we identified the main automorphism (\ref{main_automorphism_def}), (\ref{omega_transforms}) with $PT$. If we accept that $T$ changes sign to timelike $\mygen_{2 i}$ than reversion (\ref{tau_transforms}) and conjugation (\ref{conjugation_transforms}) restricted to $V$ correspond respectively to $T$ and $P$ but other identifications are possible.
\opt{margin_notes}{\mynote{mbh.note: here we comply to referee's comment, p. 648}}%
A word of caution on this point: when we deal with complex representation of real algebras, how it is customary to do for Dirac spinors, a Wick rotation can easily swap timelike and spacelike vectors, \eg going from $\R^{3,1}$ to $\R^{1,3}$. Things would be different for real Clifford algebras but in this case our formalism take us to consider neutral spaces, $\R^{m,m}$ and again the identification of timelike and spacelike coordinates is ambiguous. On top of that there is the fact that the automorphism group of $\{\Identity_V, P, T, PT\} \myisom \Z_2 \otimes \Z_2$ is the group of permutations of $\{P, T, PT\}$ that thus can be freely permuted, and so there are no indications neither from this side. Anyhow whatever the precise identification of $P$ and $T$ their corresponding automorphisms both move also the spinorial space supporting the regular representation of \myClg{}{}{g} both sending $\Psi_{a b}$ to $\Psi_{-a -b}$ and in any case $b \ne -b$.

It is evocative to write the general form of these elements in EFB
\begin{eqnarray*}
\Identity & = & \anticomm{q_1}{p_1} \anticomm{q_2}{p_2} \cdots \anticomm{q_m}{p_m} \\
\omega & = & \comm{q_1}{p_1} \comm{q_2}{p_2} \cdots \comm{q_m}{p_m} \\
\tau & = & \left(p_1 + s \, q_1\right) \left(p_2 + s \, q_2\right) \cdots \left(p_m + s \, q_m\right) \qquad s = (-1)^{m + 1}\\
\omega \tau & = & (-1)^{m} \left(p_1 - s \, q_1\right) \left(p_2 - s \, q_2\right) \cdots \left(p_m - s \, q_m\right)
\end{eqnarray*}
and the first two result: even under the main automorphism, they do not move spinor spaces and form a group isomorphic to $\Z_2$ that provides \myClg{}{}{g} grading. The last two, have parity $(-1)^{m}$ under the main automorphism, move spinor spaces and form, together with the first two, the group of discrete automorphisms isomorphic to $\Z_2 \otimes \Z_2$ group. Moreover with (\ref{chirality_parity}) and observing that $\epsilon(-x) = (-1)^m \epsilon(x)$ we find
\begin{equation*}
\begin{array}{l l l}
\h_eigen(\Psi_{-a -b}) & = & (-1)^m \h_eigen(\Psi_{a b}) \\
\g_eigen(\Psi_{-a -b}) & = & \g_eigen(\Psi_{a b})
\end{array}
\end{equation*}
showing that, when $m$ is odd, the chirality is reversed by automorphisms that move spinor spaces; a subtler study is needed to generalize this result from neutral spaces to real spaces of different signature.

To investigate how these inner automorphisms behave on generic spinors (\ref{Fock_basis_expansion}) and not only on EFB elements we give a simple result:%
\opt{margin_notes}{\mynote{mbh.ref: see p. 443.18 and \cite[Theorem 2.6]{Porteous_1995} p. 12}}%
\begin{MS_Proposition}
\label{automorphism_on_S}
for any inner automorphism $\alpha \in C_g^{\star}$ the image of a MLI is a MLI, moreover $x S_{hg} x^{-1} = S_{hg}$ if and only if $S_{hg} x^{-1} = S_{hg}$.
\end{MS_Proposition}

\begin{proof}
The first part follows immediately from the fact that a MLI must have rank 1; for the second part let $x S_{hg} x^{-1} = S_{hg}$, then since $S_{hg}$ is a MLI we have also $S_{hg} x^{-1} = S_{hg}$; vice versa let $S_{hg} x^{-1} = S_{hg}$, since $S_{hg}$ is a MLI $x S_{hg} = S_{hg}$ and thus $x S_{hg} x^{-1} = S_{hg}$.
\end{proof}
\noindent Thus since for any spinor $\varphi = \varphi \Identity$ the identity do not change spinor space. Going to $PT$ by (\ref{Fock_basis_expansion}), (\ref{EFB_are_left_Weyl}) and (\ref{chirality_parity}) and remembering that $\omega^{-1} = \omega^3$
$$
\varphi \omega^{-1} = \omega^2 \sum_{a} \xi_{a \myeee} \Psi_{a \myeee} \omega = \omega^2 \sum_{a} \xi_{a \myeee} \epsilon(\myeee) \Psi_{a \myeee} = \omega^2 \epsilon(\myeee) \varphi = \pm \varphi
$$
and so both $\Identity_V$ and $PT$ behave as expected also on generic spinors of any $S_\myeee$. The effect of reversion automorphism (\ref{tau_transforms}) on a generic spinor is
$$
\tau \varphi \tau^{-1} = \sum_{a} \xi_{a \myeee} \Psi_{- a - \myeee}
$$
and if $\varphi$ has a defined chirality, $\omega \varphi = \h_eigen \; \varphi$, then with (\ref{Weyl_eigenvector}) and (\ref{chirality_parity})
$$
\omega \tau \varphi \tau^{-1} = \sum_{a} \epsilon(-a) \xi_{a \myeee} \Psi_{- a - \myeee} = (-1)^m \h_eigen \tau \varphi \tau^{-1} \dotinformula
$$

We consider now the solutions of equations like $v \varphi = 0$, where $v \in V$ and $\varphi \in S$. We observe that they must remain the same under any injective map and thus $x v \varphi x^{-1} = 0$ if and only if $v \varphi = 0$ and thus $\varphi x^{-1} = 0$ only if $\varphi = 0$. As it was intuitive, inner automorphisms do not change solutions of $v \varphi = 0$; in particular the solutions of $x v \varphi = 0$ are identical to those of $x v \varphi x^{-1} = 0$.

We conclude with a first characterization of transformations that stabilize spinor spaces:
\begin{MS_Proposition}
\label{automorphism_on_S3}
The automorphims such that $x S_{\myeee} x^{-1} = S_{\myeee}$ form a subgroup of $C_g^{\star}$ (\ref{Cg_star_def}).
\end{MS_Proposition}
\begin{proof}
Let $x \in C_{\myeee} := \{x \in C_g^{\star} : x S_{\myeee} x^{-1} = S_{\myeee} \}$, by previous proposition we have $S_{\myeee} x^{-1} = S_{\myeee}$ and right multiplying by $x$ we get $S_{\myeee} = S_{\myeee} x$ with which we prove that also $x^{-1} S_{\myeee} x = S_{\myeee}$ \ie that also $x^{-1} \in C_{\myeee}$. Let $x, y \in C_{\myeee}$ then $x y S_{\myeee} y^{-1} x^{-1} = S_{\myeee}$ thus also $x y \in C_{\myeee}$.
\end{proof}

We remark that in general if $x$ leaves invariant spinor space $S_{\myeee}$ nothing can be said on its properties on a different $S_{\myeee'}$, moreover this subgroup is not normal as it is simple to see. This group will be pursued in a forthcoming companion paper.

\section{Conclusions}
\label{Conclusions}
We have proved that all orthogonal transformations of an even dimensional vector space $V$ can be seen as the restrictions of inner automorphisms of \myClg{}{}{g}. We are thus allowed to assume that also a spinor $\varphi$ must transform as $x \varphi x^{-1}$ and that, in some cases like \eg $P$ and $T$, this has the effect of moving spinor $\varphi \in S_{\myeee}$ to $x \varphi x^{-1} \in S_{- \myeee}$. This has no influence on the solutions of equations like $v \varphi = 0$ but the moved spinor $x \varphi x^{-1}$ may have opposite chirality.

The perspectives appear interesting but many things remain to be done to complete this study, one for all the classification of continuous transformations of $V$ since it is a simple exercise to verify that whereas all automorphisms generated by an odd number of generators, like \eg $\varphi \to \mygen_i \varphi \mygen_i^{-1}$, move the spinor space of $\varphi$, automorphisms where the generators appear in couples, like \eg $\varphi \to (\mygen_{2 i - 1} \mygen_{2 i}) \varphi (\mygen_{2 i - 1} \mygen_{2 i})^{-1}$, do not move the spinor space of $\varphi$. This and other issues on the group of trasformtions that stabilize spinors will be tackled in ongoing work.
\opt{margin_notes}{\mynote{mbh.note: $\varphi \to \varphi \mygen_{2 i - 1}^{-1}, \varphi \mygen_{2 i}^{-1}$ change sign of $g_i$ signature, while $\varphi \to \varphi (\mygen_{2 i - 1} \mygen_{2 i})^{-1}$ keep $g_i$ constant...}}%

\newpage
%\bigskip

\opt{x,std,AACA}{% in all cases - but arXiv & JMP - standard BibTeX bibliography
\bibliographystyle{plain} % or plain or.... see e.g.\ http://amath.colorado.edu/documentation/LaTeX/reference/faq/bibstyles.html#styles

\bibliography{mbh}
}

\opt{arXiv,JMP}{% only for arXiv & JMP we need to include here the L-A-S-T version of file .bbl
%
%
%\begin{thebibliography}{1}

%\end{thebibliography}
%
%
}

\opt{final_notes}{
\newpage

\section*{Things to do, notes, etc.......}

\noindent General things to do:
\begin{itemize}
\item internal automorphisms for odd dimensional $V$ (see at the end);
\item extend EFB to any signature of an even dimensional $V$;
\item extend EFB to any signature of $V$;
\item include also $C$ automorphism as in \cite{Varlamov_2015};
\item match exactly $P$ and $T$ to reversion (\ref{tau_transforms}) and conjugation (\ref{conjugation_transforms}) with $V$ with given signature;
\item study group $C_{\myeee}$;
\item study group \Pin{g} in general and Lorentz transformations in particular;
\item ...
\end{itemize}

\subsection*{Old, longer, proof of Proposition~\ref{Cl_automorphisms_thm}}
\begin{proof}
In a nutshell for any non degenerate vector space $V$ over $\R$ or $\C$ its universal Clifford algebra exists and is either a full matrix algebra $M$ or the direct sum of two such algebras that are isomorphic, \ie either $\myClg{}{}{g} \myisom M$ or $\myClg{}{}{g} \myisom M \oplus M'$ with $M \myisom M'$ \cite{BudinichP_1988e}.
\opt{margin_notes}{\mynote{mbh.note: note than any automorphism of \myClg{}{}{g} it's over $\F$ but the representations may be different ! Also this could be probably extended to any field of characteristics $\ne 2$.}}%
If the vector space is even dimensional $\myClg{}{}{g} \myisom M$ and the algebra is always central simple \cite{BudinichP_1988e} and, by Skolem -- Noether theorem, all its automorphisms are inner. To prove the converse, that for all odd dimensional vector spaces there are always non inner automorphism, we start by the complex field $\C$, in this case for $V = \C^{2 m - 1}$ then $\myClg{}{}{g} \myisom \C(2^{m - 1}) \oplus \C(2^{m - 1})$ and it's easy to see that the automorphism that swaps the two parts $\alpha(z_1, z_2) = (z_2, z_1)$ (hyperbolic involution) can not be inner (while it results inner for the ``successive'' Clifford algebra over $V = \C^{2 m }$ and $\myClg{}{}{g} \myisom \C(2^{m})$).
\opt{margin_notes}{\mynote{mbh.note: hyperbolic involution p. 586.}}%

In the real case $V = \R^{p, q}$ the same argument applies to $p-q \equiv 1, 5 \pmod{8}$ when, respectively, $\myClg{}{}{g} \myisom \R(r) \oplus \R(r)$ and $\myClg{}{}{g} \myisom \HH(r) \oplus \HH(r)$, for some $r$. It remains only the case $p-q \equiv 3, 7 \pmod{8}$ then $\myClg{}{}{g} \myisom \C(r)$ and the algebra is simple but not central since the center is $\C$. Any automorphism $\alpha$ must map the center onto itself, so the restriction of $\alpha$ to $\C$ is necessarily an $\R$-linear automorphism of $\C$ and it is therefore either the identity or complex conjugation; if the restriction is the identity then $\alpha$ is a $\C$-linear automorphism and for Skolem -- Noether theorem it is inner. But if the restriction is complex conjugation, then composing $\alpha$ with complex conjugation yields a $\C$-linear automorphism, which must be inner. But our algebra is real and the complex field, considered as a real algebra, has no inner automorphism providing complex conjugation, so in all these cases $\alpha$ alone is not inner.
\end{proof}

\subsection*{Pieces of old Introduction}
Recently Varlamov \cite{Varlamov_2001,Varlamov_2015} have initiated the complex work of merging the \Pin{p,q} \OO{p, q} hierarchies with CPT seen as principal automorphisms of \myCl{p}{q}{} and connecting them with Dabrowski groups \cite{Dabrowski_1988} and he started showing that discrete transforms correspond to automorphisms of \myCl{p}{q}{} and have proceeded to classify automorphisms of \myCl{p}{q}{} over $\C$ and $\R$ showing that the eight double coverings (Dabrowski groups \cite{Dabrowski_1988}) correspond to the eight types of real clifford algebras (the ``spinorial clock'' of \cite{BudinichP_1988e}).
\opt{margin_notes}{\mynote{mbh.note: investigate quoted refs (matrix representations of the fundamental automorphisms of $\C^n$ was first obtained by Rashevskii in 1955) + investigate group of discrete transforms}}%
With this point of view we can study all transformations within the well defined arena of Clifford algebras; this also give more consistency to the Cartan-Dieudonné theorem since all orthogonal transformations of $V$ can be seen as ``generated'' by reflections that, in a sense, form the base of all automorphisms since it is well known that any orthogonal transformation, \ie automorphism, of $V$ lifts to an automorphism of \myCl{p}{q}{} since $V$ generates \myCl{p}{q}{}.
\opt{margin_notes}{\mynote{mbh.ref: Cartan Dieudonné theorem p. 356.2 and 579}}%

In Clifford algebra \myCl{p}{q}{} the group \Pin{p,q} is a double covering of the orthogonal group \OO{p, q} and in accordance with squares of elements of the discrete subgroup $(a = P^2, b = T^2, c = (P T)^2)$ there exist eight double coverings (Dabrowski groups \cite{Dabrowski_1988}) of the orthogonal group defining by the signatures $(a, b, c)$, where $a, b, c \in \{-, +\}$. Such in brief is a standard description scheme of the discrete transformations.
\opt{margin_notes}{\mynote{mbh.note: for a neat description of Dabrowski's results check Phys\_666 p. 9}}%

But traditionally whereas \Pin{p,q} is completely built within a Clifford algebra \myCl{p}{q}{} the discrete subgroup is introduced in the standard scheme in an external way.

\subsection*{Old parts of section 5}
We have seen that spinors are MLI and that the $2^m$ spinor spaces are identified by their $h \circ g$ signatures. We state now a few useful propositions:%
\opt{margin_notes}{\mynote{mbh.note: see p. 443.18}}%
\begin{MS_Proposition}
\label{automorphism_on_S_old}
for any inner automorphism $\alpha \in C_g^{\star}$ the image of a MLI is a MLI, namely $\alpha(S_{{hg}_1}) = S_{{hg}_2}$ or $x S_{{hg}_1} x^{-1} = S_{{hg}_2}$.
\end{MS_Proposition}

\begin{proof}
Since $S$ is a MLI we have $aS = S$ so let's call $T = x S x^{-1}$ and let's consider $\mu T$. We can always write $\mu = x \nu x^{-1}$ then $\mu T = x \nu x^{-1} x S x^{-1} = x \nu S x^{-1} = x S x^{-1} = T$ thus T is a LI. Let's suppose now that T is not minimal \ie that there exists a proper subset $T' \subset T$ such that $\mu T' = T'$ for any $\mu \in \myClg{}{}{g}$. Since the automorphism is bijective there must exist a proper subset $S' \subset S$ that can be obtained by $S' = x^{-1} T' x$ and, with the same procedure of the first part of this proof, we can easily derive that also $S'$ is a LI strictly contained in $S$ but these violates the hypothesis that $S$ is a MLI so we can conclude that $x S x^{-1}$ is a MLI.
\end{proof}
It remains to be established if it is the same MLI or not. Since $S_{hg}$ is a MLI it follows $x S_{hg} = S_{hg}$ and so the only possibility that the $h \circ g$ signature of $S_{hg}$ is modified is from $S_{hg} x^{-1}$; this is proved by:
\begin{MS_Proposition}
\label{automorphism_on_S2_old}
$x S_{hg} x^{-1} = S_{hg}$ if and only if $S_{hg} x^{-1} = S_{hg}$.
\end{MS_Proposition}

\begin{proof}
Let $x S_{hg} x^{-1} = S_{hg}$, then since $S_{hg}$ is a MLI we have also $S_{hg} x^{-1} = S_{hg}$. Vice versa let $S_{hg} x^{-1} = S_{hg}$, since $S_{hg}$ is a MLI we can left multiply the first term by $x$ to get $x S_{hg} x^{-1} = S_{hg}$.
\end{proof}

\bigskip

To better characterize the automorphisms of $C_{\myeee}$ we start observing
\begin{MS_Proposition}
\label{MLI_with_idempotent}
Every MLI contains a primitive idempotent
\opt{margin_notes}{\mynote{mbh.note: I have already seen this proposition on Wiki: see p. 548.9.6 that contains very similar arguments}}%
\end{MS_Proposition}
\begin{proof}
Any MLI contains EFB elements with all possible $2^m$ $h$ signatures and the same $h \circ g$ signature, consequently any MLI contain one EFB element whose $g$ signature is $- \myeee$ \ie an element of the abelian subalgebra of \myCl{m}{m}{} that is idempotent; in EFB formalism it is the element $\Psi_{b b}$ and $\Psi_{b b}^2 = \Psi_{b b}$. Since the subalgebra formed by this element alone is a division algebra it follows that the idempotent is primitive.
\end{proof}

We can use now the standard Peirce decomposition for the algebra \myCl{m}{m}{}, indicating with $p$ the idempotent and $\Cl$ for our algebra:
$$
\Cl = p \, \Cl \, p + (1 - p) \, \Cl \, p + p \, \Cl \, (1 - p) + (1 - p) \, \Cl \, (1 - p)
$$
where $1 - p$ is the non primitive idempotent orthogonal to $p$ since $p (1 - p) = 0$. If $p$ is the primitive idempotent of the MLI $S$, since $\Cl \, p$ is a MLI it follows $S = \Cl \, p$.

An element of $C_x$ is such that $S x^{-1} = S$ or, expressed in the forms of the Peirce decomposition
\begin{eqnarray*}
\Cl \, p \left[ p \, \Cl \, p + (1 - p) \, \Cl \, p + p \, \Cl \, (1 - p) + (1 - p) \, \Cl \, (1 - p) \right] = \\
= \Cl \, p \, \Cl \, p + \Cl \, p \, \Cl \, (1 - p) \overset{?}{=} \Cl \, p
\end{eqnarray*}
that, since $\Cl \, p \, \Cl \, p = \Cl \, p$, to be satisfied requires that the form of the automorphisms of $C_x$ is:
$$
x^{-1} = p \, \Cl \, p + (1 - p) \, \Cl
$$
since $\Cl \, p x^{-1} = \Cl \, p$
and it is easy to prove that also $x$ must have this form. So the elements of \myCl{m}{m}{} that do not change the MLI $S$ that contains the primitive idempotent $p$ are those without the terms $p \, \Cl \, (1 - p)$.

In EFB language Peirce decomposition becomes: $p = \Psi_{b b}$ and $1 - p = \Identity - \Psi_{b b}$ and so the form of the elements $x$ that generate automorphism that leave invariant the MLI $S_b$ that supports the regular representation is:
$$
x^{-1} = \Psi_{b b} + (1 - \Psi_{b b}) \, \Cl
$$
that means that the generic element has the form $x^{-1} = \xi_{b b} \Psi_{b b} + \sum_{a c; a \ne b} \xi_{a c} \Psi_{a c}$ so that
$$
\Psi_{a b} x^{-1} = \Psi_{a b} \left(\xi_{b b} \Psi_{b b} + \sum_{a c; a \ne b} \xi_{a c} \Psi_{a c}\right) = \xi_{b b} \Psi_{a b} \qquad \forall a \dotinformula
$$
Of the four fundamental automorphism discussed in section~\ref{Automorphisms} only the identity and the main automorphism fulfill this condition.

\subsection*{Old conclusions}
We conclude observing that even if the general transformation of a spinor under an algebra isomorphism is $x \varphi x^{-1}$ there are cases in which this general form is irrelevant. The typical case is when we are interested in the null vector $v \in V$ associated to a given spinor, \ie to the solution of the equation $v \varphi = 0$ where $\varphi \in S$ as are for example the solutions of the Weyl's equation. Under a general automorphism $x$ the relation transforms to
$$
x v x^{-1} x \varphi x^{-1} = x v \varphi x^{-1} = 0
$$
and it is clear that the solution are the same as those obtained transforming the spinor with just $x \varphi$ and forgetting of the possible change of the MLI. So in this case the traditional transformation of spinors, even if wrong, allows to get the correct solutions. In other words what matters for the solution of the Weyl equation is just the $h$ signature of $\varphi$ that is the same in \emph{all spinor spaces} $S_{hg}$ for any of the possible $2^m$ values of $h \circ g$.

The other observation is that this could be a fully general mechanism through which the multiple spinor spaces $S_{hg}$, recently proposed for mirror particles \cite{Pavsic_2010}, are populated.

\subsection*{Main Automorphism in odd dimensional spaces}
In odd dimensional spaces $n = 2 m + 1$ then $\omega$ belongs to the center of the algebra that is thus not central. In general the main automorphism (\ref{main_automorphism_def}) is \emph{not inner} since it sends $\omega \to -\omega$ but since $\omega$ belongs to the center $\forall x \in \myClg{}{}{g}$ then $x \omega x^{-1} = \omega$. At least for the cases $\myClg{}{}{g} \myisom M \oplus M'$, it probably can be represented with a swap matrix. See log. p. 663; Spinorial Chessboard pp. 69, 73; moreover for $\myCl{0}{1}{} \myisom \C$ we know it's impossible.

In B\&T \cite[p.43]{Benn_1987} there is a proof that for $n$ odd any orthogonal automorphism that leaves the volume element $\omega$ invariant is inner; for us this applies either to reversion or to conjugation but not to both since, necessarily, one has to be even and the other has to be odd. Still this proposition probably do not contradict Varlamov \cite{Varlamov_2001,Varlamov_2015} since there they use representations.

} % note finali: stampate solo se all'inizio c'è l'opzione final_notes

\end{document}